\documentclass[aip,preprint]{revtex4}%
\usepackage[T1]{fontenc}
\usepackage[latin9]{inputenc}
\usepackage{amsmath}
\usepackage{amssymb}
\usepackage{babel}
\usepackage{amsfonts}
\usepackage{graphicx}
\usepackage{float}
\usepackage{placeins}

\usepackage[colorlinks=true, citecolor = blue,  linkcolor=black, urlcolor =
magenta, filecolor = orange]{hyperref}%

\begin{document}
\title{Relativistic spin-0 particle in a box: bound states, wavepackets, and the
disappearance of the Klein paradox}
\author{M.\ Alkhateeb}
\email{mohammed.alkhateeb@cyu.fr}
\author{A. Matzkin}
\email{alexandre.matzkin@cnrs.fr}
\affiliation{Laboratoire de Physique Théorique et Modélisation, CNRS Unité 8089, CY Cergy
Paris Université, 95302 Cergy-Pontoise cedex, France}

\begin{abstract}
The \textquotedblleft particle-in-a-box\textquotedblright\ problem is
investigated for a relativistic particle obeying the Klein-Gordon equation. To
find the bound states, the standard methods known from elementary
non-relativistic quantum mechanics can only be employed for ``shallow''
wells.\ For deeper wells, when the confining potentials become supercritical,
we show that a method based on a scattering expansion accounts for Klein
tunneling (undamped propagation outside the well) and the Klein paradox
(charge density increase inside the well). We will see that in the infinite
well limit, the wavefunction outside the well vanishes and Klein tunneling is
suppressed: quantization is thus recovered, similarly to the non-relativistic
particle in a box. In addition, we show how wavepackets can be constructed
semi-analytically from the scattering expansion, accounting for the dynamics
of Klein tunneling in a physically intuitive way.

\end{abstract}
\maketitle

\section{Introduction}

In non-relativistic quantum mechanics, the \textquotedblleft particle in a
box\textquotedblright, i.e. when the square well potential is extended to
infinite depth, is the simplest problem considered in textbooks, usually in
order to introduce the quantization of energy levels. In contrast, in the
first quantized relativistic quantum mechanics (RQM), the situation is not so
simple, and the problem is understandably hardly treated in RQM textbooks. The
reason is that when the potential reaches a sufficiently high value, the
energy gap $2mc^{2}$ separating the positive energy solutions from the
negative energy ones is crossed ($m$ is the rest mass of the particle). For
such potentials, known as \textquotedblleft supercritical
potentials\textquotedblright, the wave function does not vanish outside the
well but propagates undamped in the high potential region, a phenomenon known
as Klein tunneling \cite{dombey,wachter}. Indeed, RQM -- although remaining a
single-particle formalism -- intrinsically describes a generic quantum state
as a superposition of positive energy solutions (related to particles) and
negative energy solutions (related to antiparticles).\ 

Therefore, for relativistic particles, the particle-in-a-box problem is not
suited to introductory courses. For this reason, only finite,
non-supercritical rectangular potential wells are usually presented in
RQM\ classes (see for example Sec.\ 9.1 of Ref. \cite{strange} for the Dirac
equation describing fermions in a square well, or Sec.\ 1.11 of the textbook
\cite{greiner} for the Klein-Gordon equation, spin-0 bosons, in a radial
square well). For a Dirac particle in an infinite well, a \textquotedblleft
bag\textquotedblright\ model was developed by not introducing an external
potential, but assuming a variable mass taken to be constant and finite in a
box, but infinite outside \cite{alberto1,vidal}; in this way Klein tunneling
is suppressed and solutions similar to those known in the non-relativistic
case can be obtained. This method was recently extended to the Klein-Gordon
equation \cite{alberto2}.

In this work, we show that for the Klein-Gordon equation in a one-dimensional
box, it is not necessary to change the mass to infinity outside the well in
order to confine the particle. To do so, we shall consider multiple scattering
expansions inside the well.\ Such expansions were recently employed to
investigate relativistic dynamics across supercritical barriers
%\cite{scirep,paper2}.
\cite{paper2}. We will see below that Klein tunneling, which is prominent for
a supercritical potential well sufficiently higher than the particle energy
placed inside, disappears as the well's depth $V$ is increased. In the
infinite-well limit, Klein tunneling is suppressed and the walls of the well
become perfectly reflective, as in the non-relativistic case.

The relativistic bosonic particle in a box is an interesting problem because
it yields a simple understanding, in the first quantized framework, of the
charge creation property that is built into the Klein-Gordon equation,
extending tools (scattering solutions to simple potentials) usually
encountered in introductory non-relativistic classes. Moreover, as we will
show in this paper, time-dependent wavepackets can be easily built from the
scattering solutions.\ This is important because wavepackets allow us to
follow in an intuitive way the dynamics of charge creation in a relativistic
setting. The physics of charge creation in the presence of supercritical
potentials is much more transparent for the Klein-Gordon equation than for the
Dirac equation, which needs to rely in the first quantized formulation on hole
theory (see \cite{nitta} for a Dirac wavepacket approach for scattering on a
supercritical step).

The paper is organized as follows. We first recall in Sec. \ref{kgeq} the
Klein-Gordon equation and address the finite square well problem, obtaining
the bound-state solutions. In Sec. \ref{mse} we introduce the method of the
multiple scattering expansion (MSE) in order to calculate the wave function
inside and outside a square well. We will then see (Sec. IV) that the
wavefunction outside the well vanishes as the well depth tends to infinity.
The fixed energy solutions are similar to the well-known Schrödinger ones.
Finally, we show (Sec \ref{wpdyna}) how the MSE can be used to construct
simple wavepackets in a semi-analytical form. We will give illustrations
showing the time evolution of a Gaussian initially inside square wells of
different depths.

\section{Klein-Gordon solutions for a particle in a square well}

\label{kgeq}

\subsection{The Klein-Gordon equation}

The wavefunction $\Psi(t,x)$ describing relativistic spin-0 particles is
well-known to be described by the Klein-Gordon (KG) equation
\cite{greiner,strange}. In one spatial dimension and in the presence of an
electrostatic potential energy $V(x)$, the KG equation is expressed in the
canonical form and in the minimal coupling scheme as:
\begin{equation}
\lbrack i\hbar\partial_{t}-V(x)]^{2}\Psi(t,x)=(c^{2}\hat{p}^{2}+m^{2}%
c^{4})\Psi(t,x) \label{kg}%
\end{equation}
where $c$ is the speed of light in vacuum, $\hat{p}=-i\hbar\partial_{x}$ is
the momentum operator and $\hbar$ is the reduced Planck constant. The charge
density $\rho(t,x)$, which can take positive or negative values associated
with particles and anti-particles, is given by (see e.g. Refs.
\cite{greiner,strange})
\begin{equation}
\rho(t,x)=\frac{i\hbar}{2mc^{2}}[\Psi^{\ast}(t,x)\partial_{t}\Psi
(t,x)-\Psi(t,x)\partial_{t}\Psi^{\ast}(t,x)]-\frac{V(x)}{mc^{2}}\Psi^{\ast
}(t,x)\Psi(t,x). \label{charge}%
\end{equation}
A generic state may contain both particle and anti-particle contributions,
corresponding to positive and negative energies respectively [see Eq.
(\ref{een}) below]. The scalar product of two wave functions $\Psi_{I}(t,x)$
and $\Psi_{II}(t,x)$ is defined as:
\begin{equation}%
\begin{split}
<\Psi_{I}(t,x)|\Psi_{II}(t,x)>=\int dx\{  &  \frac{i\hbar}{2mc^{2}}[\Psi
_{I}^{\ast}(t,x)\partial_{t}\Psi_{II}^{{}}(t,x)-\partial_{t}\Psi_{I}^{\ast
}(t,x)\Psi_{II}(t,x)]\\
-  &  \frac{V(x)}{mc^{2}}[\Psi_{I}^{\ast}(t,x)\Psi_{II}(t,x)]\}.
\end{split}
\end{equation}

\subsection{The finite square well}

\subsubsection{Plane-wave solutions\label{pws}}

\begin{figure}[htb]
	\includegraphics[scale=0.25]{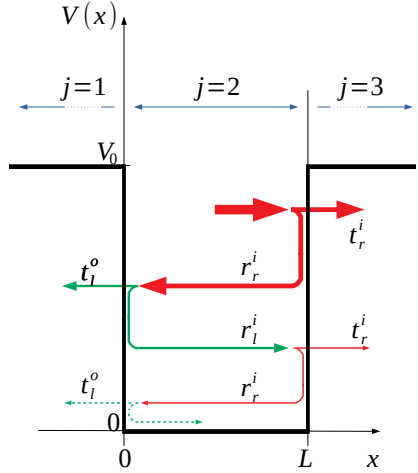} \caption{A square well with
		the 3 regions $j$ considered in the text. The arrows depict the multiple
		scattering expansion for a wave initially traveling toward the right edge of
		the well (see Sec. \ref{mse} for details).}%
	\label{figure1}%
\end{figure}

Before getting to the problem of a particle in an infinite well, let us
address first a particle inside a square well of finite depth. A square well
in one dimension can be described by the potential.
\begin{equation}
V(x)=V_{0}\theta(-x)\theta(x-L) \label{pot}%
\end{equation}
where $\theta(x)$ is the Heaviside step function, $V_{0}$ is the depth of the
well and $L$ is its width. As illustrated in Fig. \ref{figure1}, we consider
the three regions indicated by $j=1,2,3$. In each of the three regions the KG
equation (\ref{kg}) accepts plane wave solutions of the form
\begin{equation}
\Psi_{j}(t,x)=(A_{j}e^{ip_{j}x/\hbar}+B_{j}e^{-ip_{j}x/\hbar})e^{-iEt/\hbar}
\label{sol}%
\end{equation}
where we set $E$ to be the energy inside the well (region 2). By inserting
those solutions in Eq. (\ref{kg}), one obtains $E$ in terms of the momentum
inside the well
\begin{equation}
E(p)=\pm\sqrt{c^{2}p^{2}+m^{2}c^{4}} \label{een}%
\end{equation}
where for convenience we put $p=p_{2}$. These are the plane wave
solutions in free space known from RQM textbooks \cite{greiner,strange}. A plane wave with
$E(p)>0$ represents a particle, whereas a solution with $E(p)<0$ represents an antiparticle. We will be 
considering situations in which a particle is placed inside the well, so we will take positive plane-wave solutions in region 2.
Outside the well (in regions 1 and 3), it is straightforward to
see that $\Psi_{j}(t,x)$ is a solution provided $p_{1,3}=q(p)$ where
\begin{equation}
q(p)=\pm\sqrt{(E(p)-V_{0})^{2}-m^{2}c^{4}}/c. \label{qdef}%
\end{equation}

Note that in the limit of an infinite\ well ($V_{0}\gg mc^{2}$), $q(p)$ is
always real, so that typical solutions $\Psi_{j}(t,x)$ in regions $j=1,3$ are
oscillating. Note also that a classical particle with energy $E$ would have
inside the well a velocity $v=pc^{2}/E$ (so $p$ and $v$ have the same sign).
Hence for the region 2 solutions $\Psi_{2}(t,x),$ when $p$ is positive,
$e^{ipx/\hbar}$ corresponds to a particle moving to the right (and
$e^{-ipx/\hbar}$ to the left). However outside the well we have
\begin{equation}
v^{\prime}=qc^{2}/(E-V_{0}) \label{sign}%
\end{equation}
so that for large $V_{0}$ the velocity and the momentum of a classical
particle have opposite signs \cite{barut,costella}. So a plane wave
$e^{iqx/\hbar}$ with $q>0$ now corresponds to a particle moving to the left.
This can also be seen by rewriting the plane waves in terms of the energy
outside the well, say $\exp{i(p_{1}x-\bar{E}t)/\hbar}$. This is tantamount to
taking the potential to be 0 in region 1 and $\bar{V}=-V_{0}$ in region 2
(with $V_{0}>0$). Since we require $\bar{E}-\bar{V}$ to be positive and
smaller than $V_{0}$ (in order to represent a particle inside the well), we
must have $\bar{E}<0$. In this case, a given point $x$ described by the plane
wave $\exp{i(p_{1}x-\bar{E}t)/\hbar}$ travels to the right if $p_{1}$ is
negative. For instance the position of an antinode changes by $\Delta x=\Delta
t\bar{E}/p_{1}$ in the time interval $\Delta t$, so if $\bar{E}<0$, the sign
of $\Delta x$ will be opposite to the sign of $p_{1}$ \footnote{We thank an
anonymous referee for suggesting this argument.}.

\subsubsection{Bound states \label{bs}}

Bound states are obtained when the solutions outside the well are
exponentially decaying. This happens when $q(p)$ has imaginary values, that is
for potentials satisfying $E-mc^{2}<V_{0}<E+mc^{2}$.\ Note that for a particle
at rest in the well frame, $E\approx mc^{2}$ and the condition for the
existence of bound states becomes $V_{0}<2mc^{2}$.

In order to find the bound state solutions, we employ the same method used in
elementary quantum mechanics for the Schrödinger equation square well. We
first set the boundary conditions on the wavefunctions (\ref{sol}) accounting
for no particles incident from the left in region 1 nor from the right in
region 3, yielding
\begin{equation}
A_{1}=B_{3}=0. \label{boundary}%
\end{equation}
We then require the continuity of the wave functions $\Psi_{j}(t,x)$ of Eq.
(\ref{sol}) and their spatial derivatives at the potential discontinuity
points $x=0$ and $x=L$:
\begin{equation}%
\begin{split}
\Psi_{1}(t,0)  &  =\Psi_{2}(t,0),\quad\Psi_{2}(t,L)=\Psi_{3}(t,L)\\
\Psi_{1}^{\prime}(t,0)  &  =\Psi_{2}^{\prime}(t,0),\quad\Psi_{2}^{\prime
}(t,L)=\Psi_{3}^{\prime}(t,L).
\end{split}
\label{match}%
\end{equation}
This gives
\begin{equation}%
\begin{split}
B_{1}=A_{2}+B_{2},\quad &  A_{2}e^{ipL}+B_{2}e^{-ipL}=A_{3}e^{iqL}\\
-qB_{1}=p(A_{2}-B_{2}),\quad &  p(A_{2}e^{ipL}-B_{2}e^{-ipL})=qA_{3}e^{iqL}.\\
&
\end{split}
\label{matchingcond}%
\end{equation}
By eliminating $A_{3}$ and $B_{1}$ we obtain a system of two equations in
$A_{2}$ and $B_{2}$
\begin{equation}%
\begin{split}
(q+p)A_{2}+(q-p)B_{2}  &  =0\\
(q-p)A_{2}e^{ipL}+(q+p)B_{2}e^{-ipL}  &  =0,\\
&
\end{split}
\label{boundeqA2B2}%
\end{equation}
where $q$ is given by Eq. (\ref{qdef}). This system admits nontrivial
solutions when the determinant of the system (\ref{boundeqA2B2}) vanishes,
\begin{equation}
(q+p)^{2}e^{-ipL}-(q-p)^{2}e^{ipL}=0. \label{det}%
\end{equation}

\begin{figure}[htb]
\includegraphics[scale=.75]{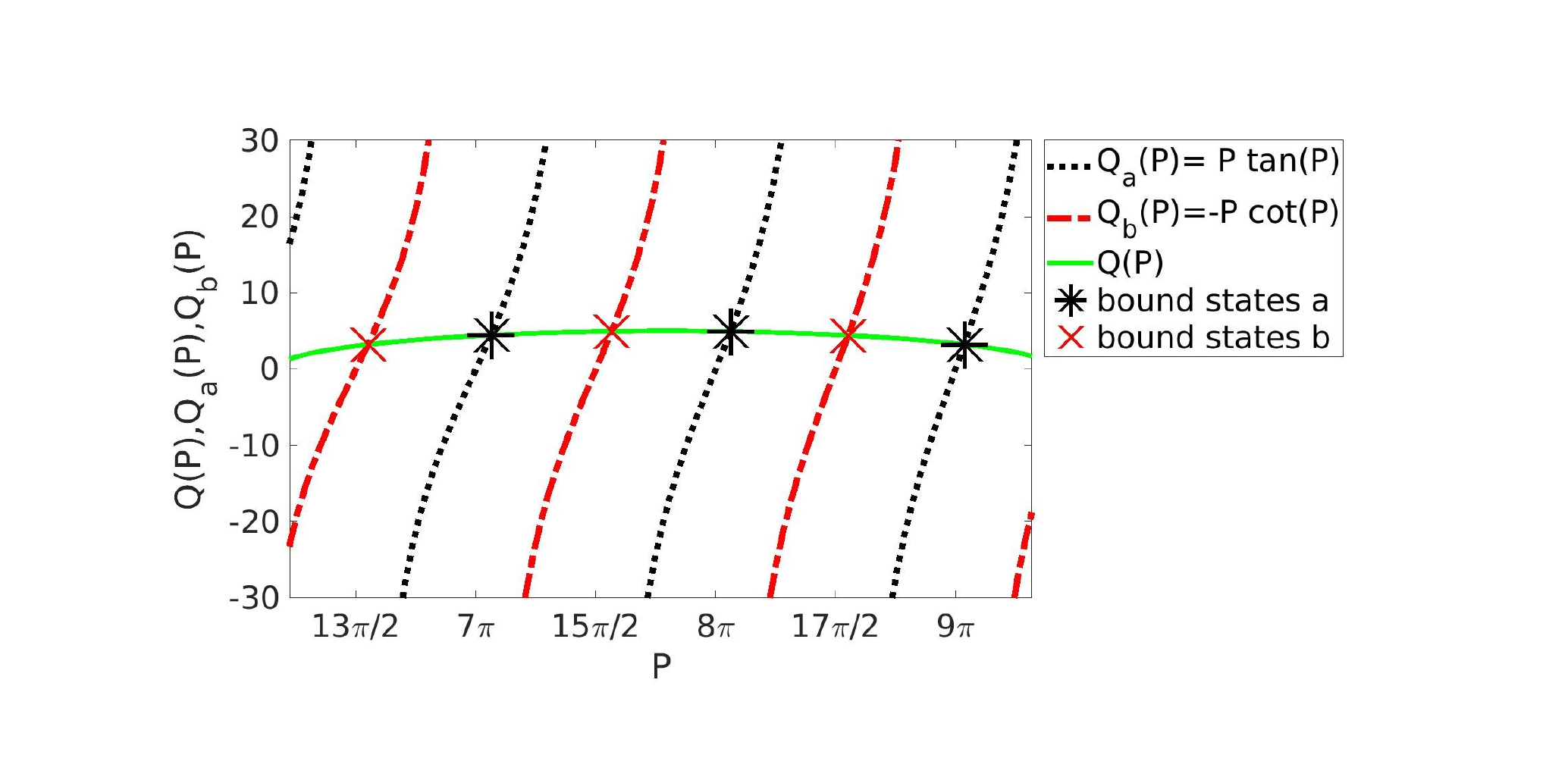} \centering
. \caption{The bound state energies of a particle of mass $m=1$ ($L=10$,
natural units are used, see text) are found from the values of $P=pL/(2\hbar)$
at the intersections of the curves defined in Eq. (\ref{split}).}%
\label{boundstatesfig}%
\end{figure}

Nontrivial solutions exist only if $q$ is an imaginary number $q=iq_{r}$ where
$q_{r}\in\mathbb{R}$. Solving Eq. (\ref{det}) for $q$ gives the two
solutions:
\begin{equation}%
\begin{split}
q_{ra}  &  =p\tan(pL/2)\\
q_{rb}  &  =-p\cot(pL/2).\\
&
\end{split}
\label{tancot}%
\end{equation}
As is familiar for the Schrödinger square well \cite{lima}, the bound state
energies are obtained from the intersections of the curves $q_{ra,b}(p)$ with
the curve $q_{r}(p)=\sqrt{m^{2}c^{4}-(E(p)-V)^{2}}/c$. For simplicity, we use
the dimensionless variables
\begin{equation}%
\begin{split}
&  Q=qL/(2\hbar)\\
&  Q_{a,b}=q_{r,a,b}L/(2\hbar)\\
&  P=pL/(2\hbar)
\end{split}
\label{split}%
\end{equation}
Fig. \ref{boundstatesfig} gives an illustration for a particle confined in a
well of width $L=10$ (we employ natural units $c=\hbar=\varepsilon_{0}=1$ as
well as $m=1$; the conversion to SI units depends on the particle's mass, eg
for a pion meson $\pi^{+}$ the mass is 139.57 MeV/c$^{2}$). The energies are
inferred from the values of $P$ at the intersection points.

\section{Multiple scattering expansion for supercritical wells}

\label{mse}

\subsection{Principle \label{prin}}

We have just seen that the method depending on matching conditions jointly at
$x=0$ and $x=L$ as per Eq. (\ref{match}) only works if $q(p)$ is imaginary,
since otherwise Eq. (\ref{det}) has no solutions. However, as is seen directly
from Eq. (\ref{qdef}), for sufficiently large $V_{0}$, $q(p)$ is real. For
this case we use a different method in which the wavefunction is seen as
resulting from a multiple scattering process on the well's edges. The well is
actually considered as being made out of two potential steps and the matching
conditions apply separately at each step.

More precisely, consider the following step potentials: a left step,
$V_{l}(x)=V_{0}\theta(-x)$ and a right step $V_{r}(x)=V_{0}\theta(x-L)$. Let
us focus on the wavefunction inside the well, whose general form is given by
$\Psi_{2}(t,x)$ [Eq. (\ref{sol})]; the boundary conditions are those given by
Eq. (\ref{boundary}), meaning no waves are incoming towards the well. Let us
first consider a plane wave $\alpha e^{ipx/\hbar}$ with amplitude $\alpha$
propagating inside the well towards the right ($p>0$; see Fig. \ref{figure1}).
On hitting the right step, this wave will be partly reflected and partly
transmitted to region 3. The part reflected inside the well will now travel
towards the left, until it hits the left step, at which point it suffers
another reflection and transmission. This multiple scattering process
continues as the reflected wave inside the well travels towards the right
edge. Similarly, we can consider a plane wave $\beta e^{-ipx/\hbar}$ of
amplitude $\beta$ initially inside the well but propagating to the left.\ This
wave hits the left step first and then scatters multiple times off the
2\ edges similarly. Multiple scattering expansions, generally employed when
several scatterers are involved, are also often used in potential scattering
problems in order to gain insight in the buildup of solutions involving many
reflections (see \cite{beam} for an application to plane-wave scattering on a
rectangular barrier).

\subsection{Determination of the amplitudes}

The coefficients giving the scattering amplitudes due to reflection and
transmission at the two steps will be denoted as $r_{l,r}$ and $t_{l,r}$
respectively, where $l\ $and $r$ indicate the left and right steps. In order
to calculate those coefficients, one has to solve the step problem separately
for each of the two steps.

The continuity of the plane wave $e^{ipx/\hbar}$ and its first spatial
derivative at the right step ($x=L$) yields the two equations
\begin{equation}
e^{ipL/\hbar}+r_{r}e^{-ipL/\hbar}=t_{r}e^{iqL/\hbar},\quad e^{ipL/\hbar}%
-r_{r}e^{-ipL/\hbar}=\frac{q}{p}t_{r}e^{iqL/\hbar}%
\end{equation}
giving
\begin{equation}
t_{r}=\frac{2p}{p+q}e^{i(p-q)L/\hbar},\quad r_{r}=\frac{p-q}{p+q}%
e^{i2pL/\hbar}. \label{right}%
\end{equation}

Similarly, in order to calculate the coefficients of reflection and
transmission suffered by a plane wave propagating inside the well towards the
left step, one uses the continuity of the plane wave and its space derivative
at $x=0$ to obtain:
\begin{equation}
t_{l}=\frac{2p}{p+q},\quad r_{l}=\frac{p-q}{p+q} \label{left}%
\end{equation}

After the plane wave reflects for the first time either on the right or left
steps, it will undergo a certain number of reflections before being finally
transmitted outside the well. Let $\alpha e^{ipx/\hbar}$ be the initial wave
inside the well moving to the right (recall $B_{3}=0$). After the first cycle
of reflections from both steps, the amplitude of the same plane wave becomes
$\alpha r_{r}r_{l},$ and $\alpha(r_{r}r_{l})^{n}$ after $n$ cycles of
successive reflections. This process is illustrated in Fig. \ref{figure1}. In
addition, an initial plane wave moving to the left (recall $A_{1}=0$), $\beta
e^{-ipx/\hbar}$ contributes, after reflecting on the left step, to the wave
moving to the right, first with amplitude $\beta r_{l},$ and then multiplied
by $(r_{r}r_{l})$ after each cycle of reflections. The amplitude of the plane
wave $e^{ipx/\hbar}$ in region 2 is the sum of these contributions, namely
$(\alpha+\beta r_{l})\sum_{n}(r_{r}r_{l})^{n}$. We can identify this term with
the amplitude $A_{2}$ in region 2, Eq. (\ref{sol}) (recall we have set
$p\equiv p_{2}$).

Along the same lines, we identify $B_{2}$ in Eq. (\ref{sol}) with the
amplitude of the term $e^{-ipx/\hbar}$ inside the well resulting from multiple
scattering, as well as $B_{1}$ in region 1 and $A_{3}$ in region 3. The result
is
\begin{equation}%
\begin{split}
&  B_{1}=t_{l}(\alpha r_{r}+\beta)\sum_{n=0}^{\infty}(r_{r}r_{l})^{n}\\
&  A_{2}=(\alpha+\beta r_{l})\sum_{n=0}^{\infty}(r_{r}r_{l})^{n}\\
&  B_{2}=(\alpha r_{r}+\beta)\sum_{n=0}^{\infty}(r_{r}r_{l})^{n}\\
&  A_{3}=t_{r}(\alpha+\beta r_{l})\sum_{n=0}^{\infty}(r_{r}r_{l})^{n}\\
&  A_{1}=B_{3}=0.
\end{split}
\label{msecoefeq}%
\end{equation}

The behavior of the series $\sum_{n\geq0}(r_{l}r_{r})^{n}$ is interesting as
it is related to charge creation. The term
\begin{equation}
|r_{l}r_{r}|=\left\vert \frac{p-q}{p+q}\right\vert ^{2} \label{grr}%
\end{equation}
can indeed be greater or smaller than 1, corresponding respectively to a
divergent or convergent series. As follows from Eq. (\ref{sign}), for a
supercritical potential $\left(  E-V_{0}\right)  <0$ so the direction of the
motion is opposite to the direction of the momentum. Hence given the boundary
conditions $A_{1}=B_{3}=0$, we see that we must set $q<0$ in order to
represent outgoing waves in regions 1 and 3 (moving in the negative and
positive directions respectively). We conclude that for supercritical wells
$|r_{l}r_{r}|>1$ and the amplitudes (\ref{msecoefeq}) diverge. The physical
meaning of a diverging series is best understood in a time-dependent picture,
as we will see in Sec. \ref{wpdyna}. The $n$th term of the series will be seen
to correspond to the $n$th time the wavepacket hits one of the edges, each hit
increasing the wave-packet's amplitude.

Note that for $q<0$, both $\left\vert r_{l}\right\vert >1$ and $\left\vert
r_{r}\right\vert >1$.\ This is an illustration of bosonic superradiance at a
supercritical potential step: for a given plane-wave incoming on the potential
step (here the left or right steps), the reflected current is higher then the
incoming one \cite{manogue,grobe-boson}. This phenomenon, that at first sight appears surprising, became known as the ``Klein paradox''.

\section{The infinite well}

As we have just seen, one of the signatures of the Klein-Gordon supercritical
well -- a feature unknown in non-relativistic wells -- is that the amplitudes
outside the well, $B_{1}$ and $A_{3}$, are not only non-zero, but grow with
time. Each time a particle hits an edge of the well, the reflected wave has a
higher amplitude, but since the total charge is conserved, antiparticles are
transmitted in zones 1\ and 3.

However, it can be seen that as the depth of the supercritical well increases,
the amplitudes of the wavefunction transmitted outside the well decrease.
Indeed, the step transmission coefficients $t_{r}$ and $t_{l}$ given by Eqs.
(\ref{right}) and (\ref{left}) are proportional to $1/V_{0}$. Hence in the
limit of infinite potentials, $V_{0}\rightarrow\infty$, the transmission
vanishes. We also see from Eqs (\ref{right}) and (\ref{left}) that
$r_{l}\rightarrow-1,$ $r_{r}\rightarrow-e^{2ipL/\hbar}$ and $\sum_{n}%
(r_{r}r_{l})^{n}$ is bounded and oscillates. Hence from Eq. (\ref{msecoefeq})
in this limit $A_{3}\rightarrow0$ and $B_{1}\rightarrow0$.\ This implies
$\psi(x=0)=\psi(x=L)=0$ and these conditions can only be obeyed provided
\begin{equation}
p=\frac{k\pi\hbar}{L} \label{quantiz}%
\end{equation}
(where $k$ is an integer);\ we also then have $B_{2}=-A_{2}$. The unnormalized
wavefunction inside the well takes the form
\begin{equation}
\Psi_{2}(t,x)=2iA_{2}\sin(\frac{k\pi}{L}x)e^{-i\frac{E_{k}t}{\hbar}},
\end{equation}
while the amplitudes outside the well obey $B_{1}\rightarrow0$ and
$A_{3}\rightarrow0$ (although for $p=k\pi\hbar/L$, $\sum_{n}(r_{r}r_{l})^{n}$
diverges). This can be seen by remarking that when Eq. (\ref{quantiz}) holds,
$r_{r}r_{l}=1,$ and $B_{1}$ can be parsed as
\begin{equation}
B_{1}=t_{l}(\alpha r_{r}+\beta)+t_{l}(\alpha r_{r}+\beta)+...
\end{equation}
Since $t_{l}\rightarrow0$ as $V_{0}\rightarrow\infty,$ the wavefunction in
region 1 vanishes in this limit. A similar argument holds for $A_{3}$.

Note however that $A_{2}$ (and $B_{2})$ become formally infinite, given that
the series $\sum_{n}^{n_{\max}}e^{2inpL/\hbar}=n_{\max}+1$ is unbounded when
Eq. (\ref{quantiz}) holds as $n_{\max}\rightarrow\infty$. Since the total
charge must be conserved (and cannot change each time $n_{\max}$ increases),
the wavefunction inside the well should be renormalized to the total charge.
Unit charge normalization corresponds to%

\begin{equation}
\Psi_{2}^{k}(t,x)=\sqrt{\frac{2}{L}}\sin(\frac{k\pi}{L}x)e^{-i\frac{E_{k}%
t}{\hbar}} \label{solinf}%
\end{equation}
with [Eqs. (\ref{een}) and (\ref{quantiz})]
\begin{equation}
E_{k}=\sqrt{\left(  \frac{k\pi\hbar}{L}\right)  ^{2}c^{2}+m^{2}c^{4}}.
\label{quantize}%
\end{equation}
In the non-relativistic limit, the kinetic energy is small relative to the
rest mass, yielding
\begin{equation}
E\approx E_{k}^{NR}=mc^{2}+\frac{k^{2}\pi^{2}\hbar^{2}}{2mL^{2}}%
\end{equation}
recovering the non-relativistic particle in a box energies (up to the rest
mass energy term). Eq. (\ref{quantize}) is the same result obtained recently
by Alberto, Das and Vagenas \cite{alberto2}, who employed a bag-model (taking
the mass to be infinite mass in regions 1 and 3) in order to ensure the
suppression of Klein tunneling.

In a real situation, neither $V_{0}$ nor the number of reflections
(corresponding to the time spent inside the well) can be infinite. Given a
finite value of $V_{0},$ a particle placed inside the well is represented by a
wavepacket that will start leaking after a certain number of internal
reflections, as we discuss in the next Section. This shows that although
quantization for infinitely deep wells looks similar to the corresponding
non-relativistic well, the mechanism is very different, as in the latter case
we have exponentially decreasing solutions that vanish immediately outside the
well, whereas in the present case we have oscillating solutions that are suppressed.

\begin{figure}[htb]
\includegraphics[scale=.64]{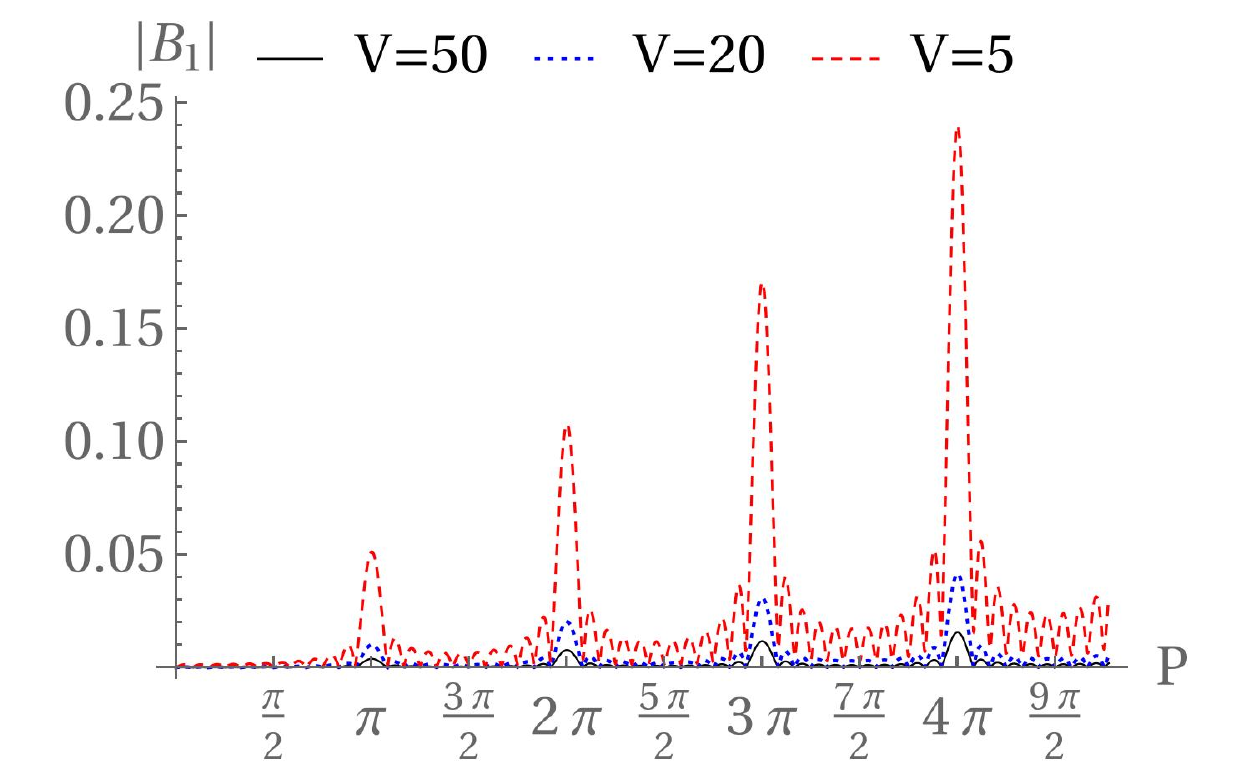} \includegraphics[scale=.64]{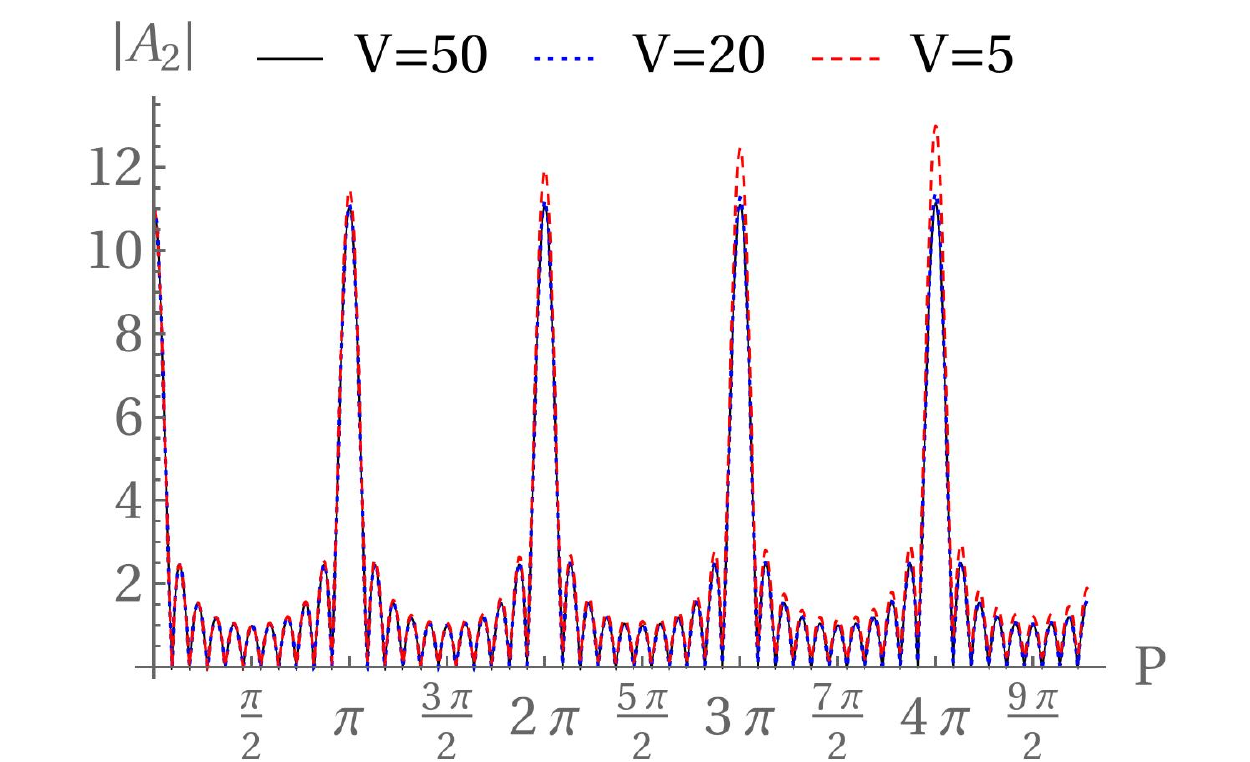}
\caption{The amplitudes $|B_{1}|$ and $|A_{2}|$ calculated using the MSE
relations Eq. (\ref{msecoefeq}) with $n_{max}=10$, $\alpha=1$ and $\beta=0$
are shown for different values of the supercritical well depth $V_{0}=5,20,50$
($p$ is given in units of $1/L$, and $c=\hbar=1$). Inside the well, the
resonant structure of $|A_{2}|$ is not much affected as $V_{0}$ changes (the
curves nearly superpose), but the amplitude $|B_{1}|$ outside the well,
indicative of Klein tunneling, is seen to decrease as $V_{0}$ increases.}%
\label{quan_coef}%
\end{figure}

Note that although quantization only appears in the limit $V_{0}%
\rightarrow\infty$, for high but finite values of $V_{0}$ resonant Klein
tunneling (e.g., Ref. \cite{barbier2008}) takes
place: the amplitudes (\ref{msecoefeq}) peak for energy values around $E_{k}$
given by Eq. (\ref{quantize}). This can be seen by plotting the amplitudes as
a function of $E$ or $p$. An illustration is given in Fig. \ref{quan_coef}
showing $B_{1}(p)$ and $A_{2}(p)$ for different values of $V_{0}$. It can be
seen that the amplitudes are peaked around the quantized $p$ values [Eq.
(\ref{quantiz})] while concomitantly decreasing as the well depth increases.

For completeness, let us mention that the square well bound states of Sec.
\ref{bs} can also be recovered employing the MSE.
%In the regimes in which the MSE diverges
%however, the joint boundary conditions (\ref{match}) become unphysical and should
%not be employed.
Indeed, for bound states, the wavefunction must be a standing wave. Given the
symmetry of the problem, the wavefunction is either symmetric or
anti-symmetric with respect to the center of the well, $x=L/2$. In the
symmetric case, the standing wave is thus given by $C \cos[ p(x - L/2)/\hbar)]
$. Matching this form to $\Psi_{2}(x) = A_{2} e^{i p x / \hbar} + B_{2} e^{-i
p x / \hbar}$ leads to
\begin{equation}
\frac{A_{2}}{B_{2}} = e^{-ipL/ \hbar}. \label{st}%
\end{equation}
The anti-symmetric standing wave is of the form $C \sin[ p(x - L/2)/\hbar)] $,
which is equated to $\Psi_{2}(x)$ to obtain ${A_{2}}/{B_{2}} = -e^{-ipL/
\hbar}$. Replacing $A_{2}$ and $B_{2}$ by their respective MSE expansion given
by Eqs. (\ref{msecoefeq}) therefore leads to
\begin{equation}
\frac{\alpha+ \beta r_{l}}{\alpha r_{r} + \beta} = \pm e^{-ipL/\hbar}.
\label{zad4}%
\end{equation}
Using $r_{r} = r_{l} e^{2ipL/ \hbar}$ from Eqs. (\ref{left})-(\ref{right}) and
keeping in mind that $\alpha$ and $\beta$ are arbitrary complex numbers, Eq.
(\ref{zad4}) becomes
\begin{equation}
r_{l} = \pm e^{-ipL}.
\end{equation}
Now, using $r_{l} = (p-q)/(p+q)$ from Eq. (\ref{left}), and squaring both
sides of this equation leads to Eq. (\ref{det}) and hence to the quantisation
conditions obtained above in Sec. \ref{bs}.

Note that these bound states are obtained when the solutions outside the well
are exponentially decaying. In this case the series $\sum_{n}\left(
r_{r}r_{l}\right)  ^{n}$ is bounded and oscillates, whereas in the
supercritical regime this series was seen to be exponentially divergent. When
the MSE diverges, applying joint matching conditions of the type given by Eq.
(\ref{match}) is incorrect and leads to unphysical results (for instance, in
the scattering of Klein-Gordon particles on a barrier, doing so leads to
acausal wavepackets and superluminal barrier traversal times
\cite{gutierrez,xu}).

\section{Wavepacket dynamics \label{wpdyna}}

\subsection{Wavepacket construction}

Since the solutions $\Psi_{j}(t,x)$ of Eq. (\ref{sol}), with the amplitudes
given by Eq. (\ref{msecoefeq}), obey the Klein Gordon equation inside and
outside the well, we can build a wavepacket by superposing plane waves of
different momenta $p$. We will follow the evolution of an initial
Gaussian-like wavefunction localized at the center of the box and launched
towards the right edge (that is with a mean momentum $p_{0}>0$). We will
consider two instances of supercritical wells, one with a \textquotedblleft
moderate\textquotedblright\ depth displaying Klein tunneling, the other with a
larger depth in which Klein tunneling is suppressed.

Let us consider an initial wavepacket
\begin{equation}
G(0,x)=\int dpg(p)(A_{2}(p)e^{ipx/\hbar}+B_{2}(p)e^{-ipx/\hbar}) \label{gau0}%
\end{equation}
with
\begin{equation}
g(p)=e^{-\frac{(p-p_{0})^{2}}{4\sigma_{p}^{2}}}e^{-ipx_{0}}%
\end{equation}
We will choose $x_{0}$ to be the center of the well and take $p_{0}$ as well
as all the momenta in the integration range in Eq. (\ref{gau0}) positive.\ We
therefore set $\beta=0$ in the amplitudes (\ref{msecoefeq}) and choose
$\alpha$ in accordance with unit normalization for the wavepacket. $\sigma
_{p}^{2}$ fixes the width of the wavepacket in momentum space (ideally narrow,
though its spread in position space should remain small relative to $L$).
Finally, the sum $\sum(r_{r}r_{l})^{n}$ is taken from $n=0$ to $n_{\max}$
where the choice of $n_{\max}$ depends on the values of $t$ for which the
wavepacket dynamics will be computed. Indeed, each term $(r_{r}r_{l})^{n}$
translates the wavepacket by a distance $2nL$, so this term will only come
into play at times of the order of $t\sim2nL/v$ where $v\sim p_{0}%
c/\sqrt{c^{2}m^{2}+p_{0}^{2}}$ is the wavepacket mean velocity. Note that in
position space $G(0,x)$ is essentially a Gaussian proportional to $e^{-\left(
x-x_{0}\right)  ^{2}/4\sigma_{x} ^{2}}e^{ip_{0}x}$ \footnote{Strictly speaking
a Gaussian in position space would have negative energy contributions not
included in $G(0,x)$ given by Eq. (\ref{gau0}). Such contributions are
negligible in the non-relativistic regime and become dominant in the
ultra-relativistic regime. For more details in the context of barrier
scattering, see \cite{paper2}.}.

Following Eq. (\ref{sol}) the wavepacket in each region is given by
\begin{equation}
G_{j}(t,x)=\int dpg(p)\Psi_{j}(t,p)\label{timeevol}%
\end{equation}
where the amplitudes $A_{j}(p)$ and $B_{j}(p)$ are obtained from the MSE. For
supercritical potential wells, we have to take $q<0$ in the MSE amplitudes.
The charge $\rho(t,x)$ associated with the wavepacket in each region is
computed from $G_{j}(t,x)$ by means of Eq. (\ref{charge}).

\begin{figure}[htb]
\label{sup} \includegraphics[scale=0.12]{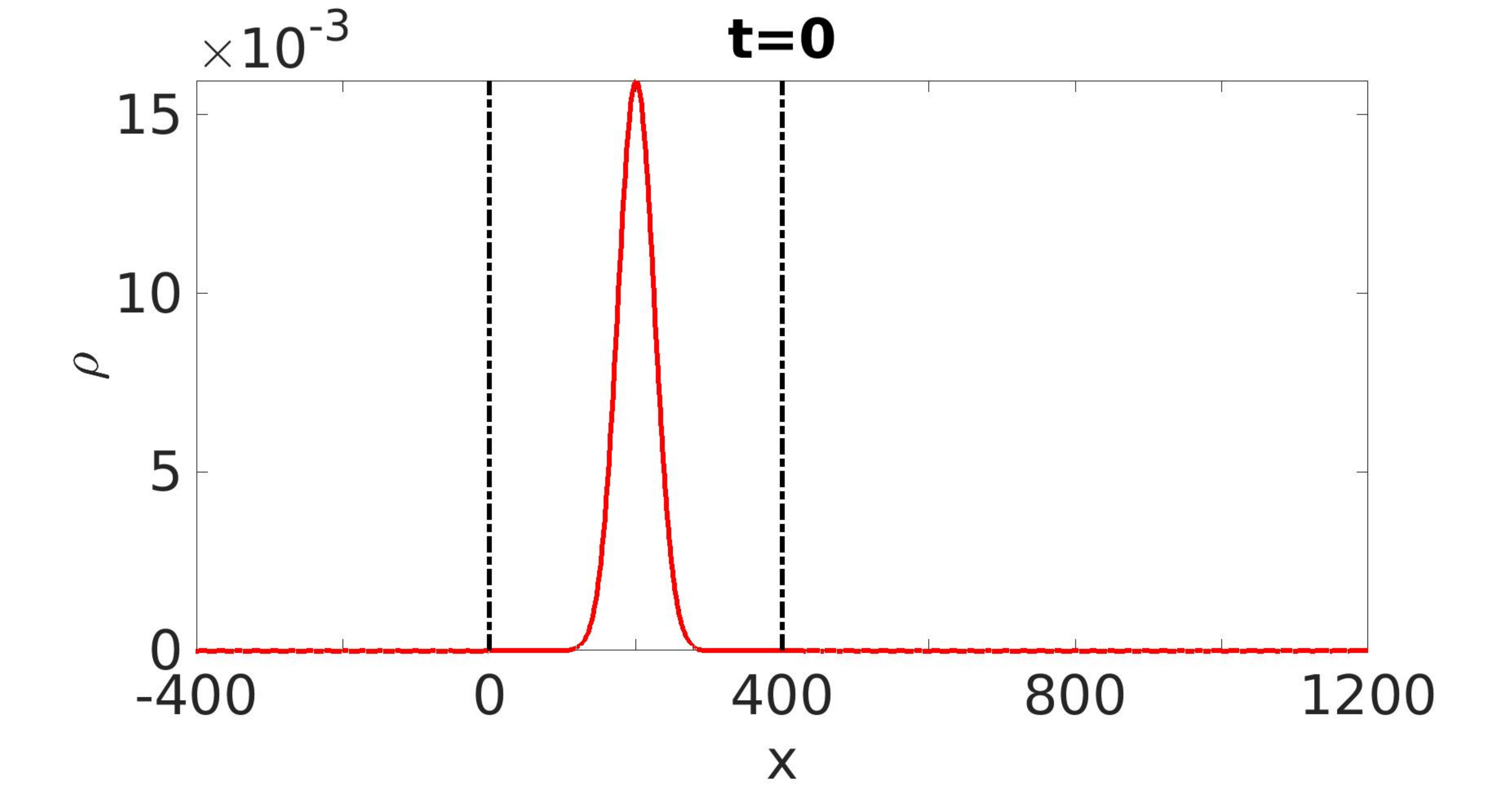}
\includegraphics[scale=0.12]{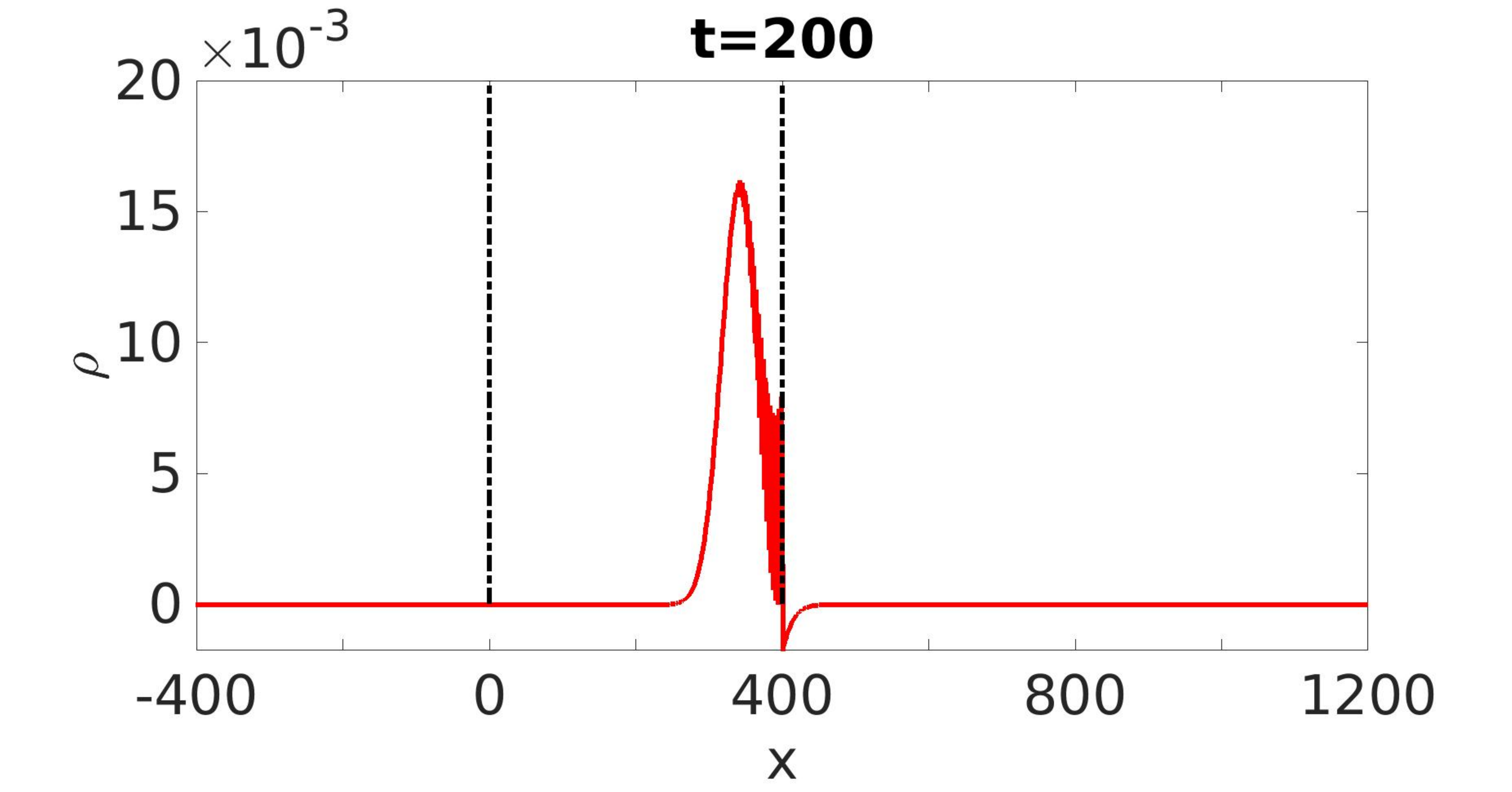}
\includegraphics[scale=0.12]{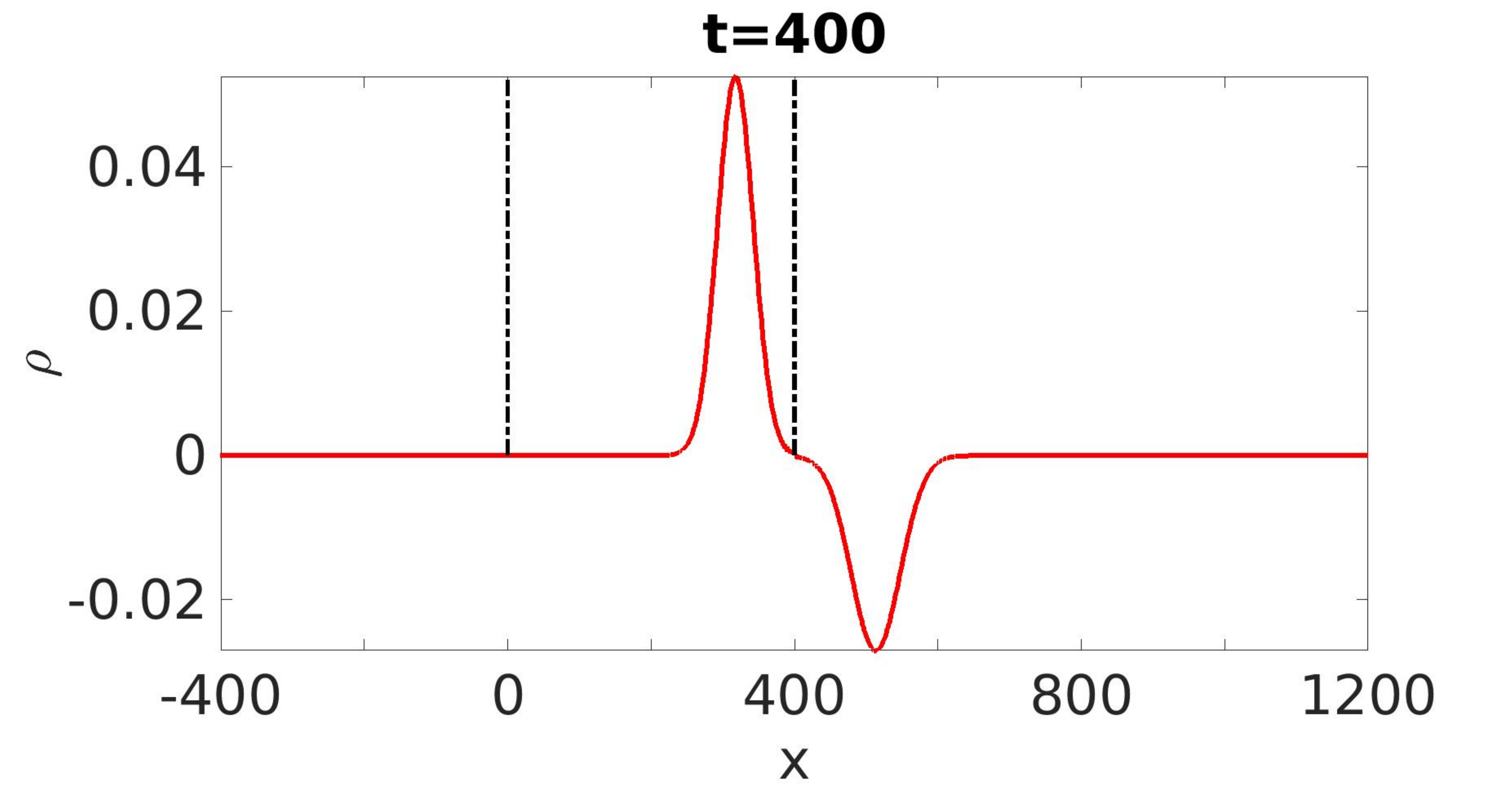}
\includegraphics[scale=0.12]{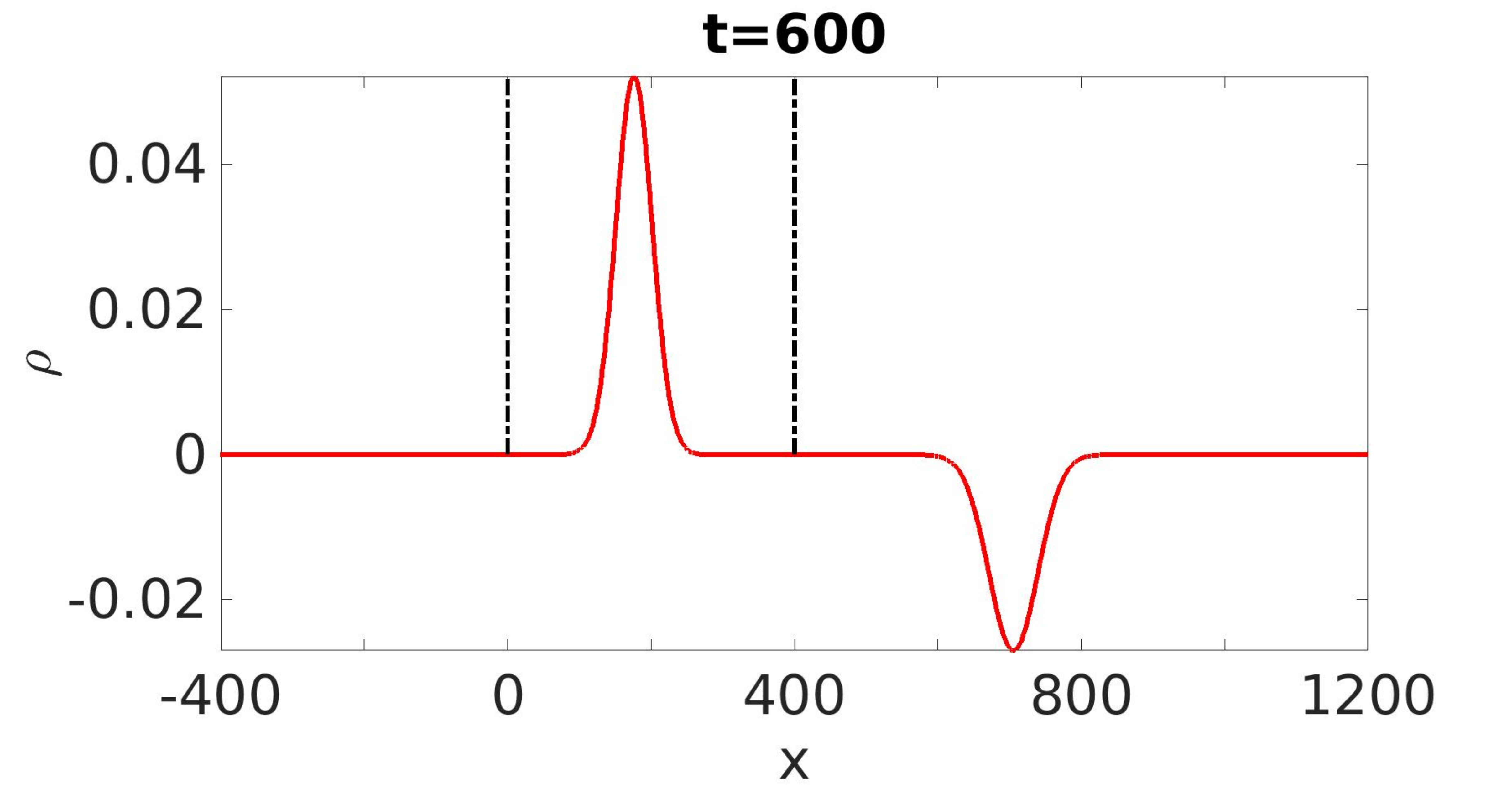}
\includegraphics[scale=0.12]{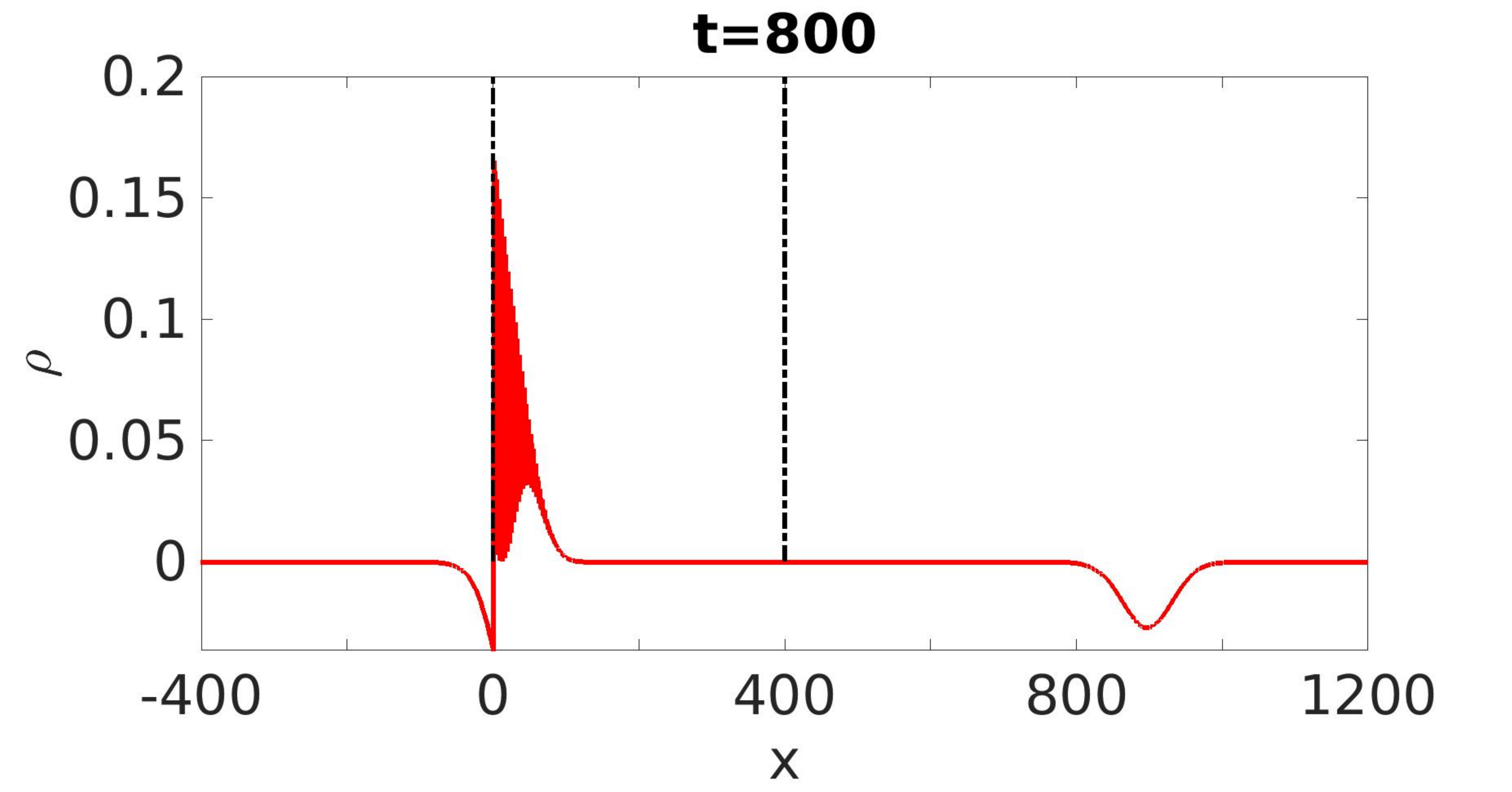}
\includegraphics[scale=0.12]{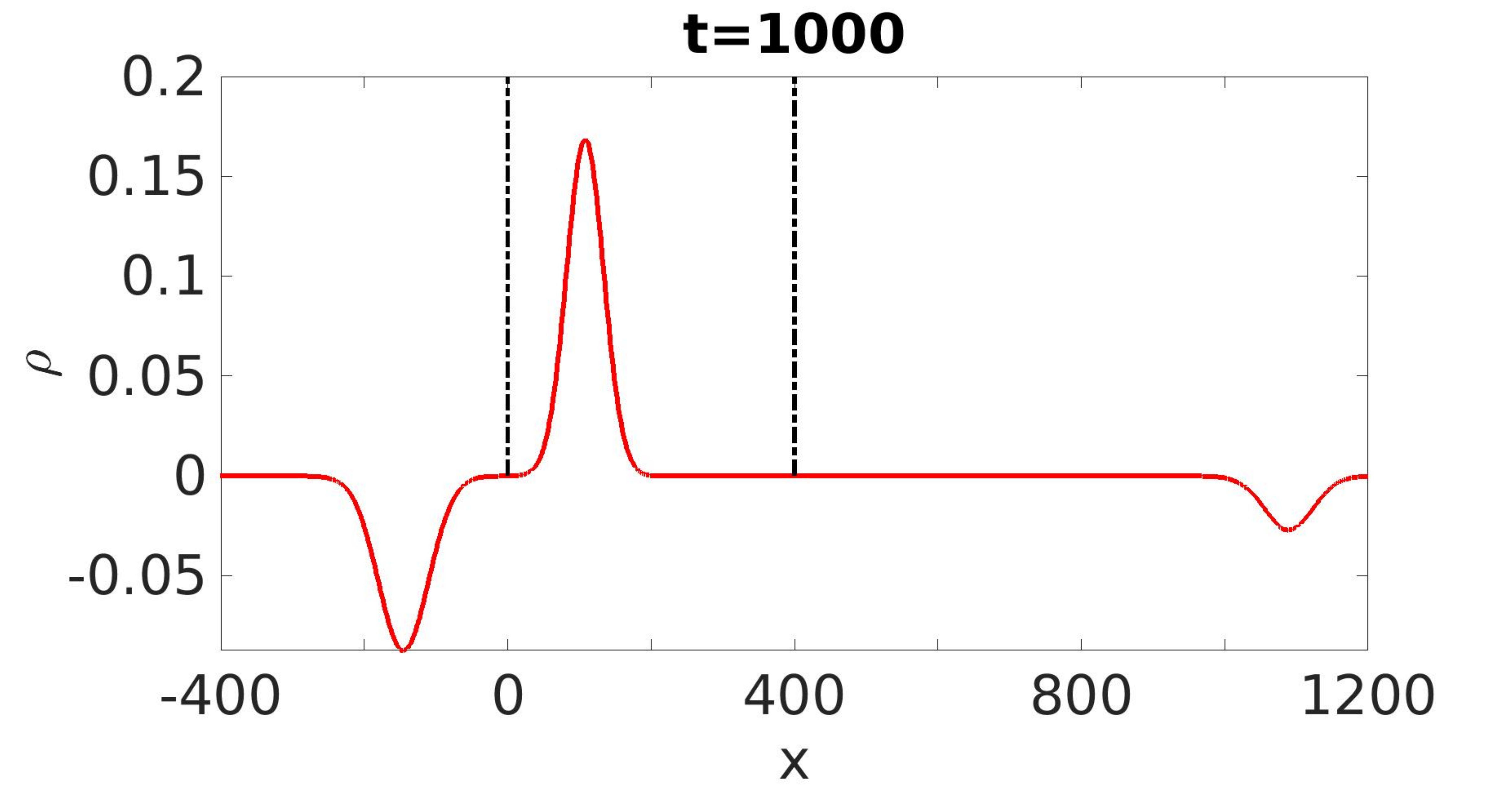} \caption{The charge $\rho(t,x)$
associated with the wavepacket given by Eq. (\ref{timeevol}) for a particle of
unit mass is shown for different times as indicated within each panel. The
parameters are the following: $L=400$ and $V_{0}=5mc^{2}$ for the well,
$x_{0}=200,p_{0}=1,\sigma_{p}=0.02$ for the initial state, $\alpha=1$,
$\beta=0$, $n_{max}=10$ for the MSE series (natural units $c=\hbar=1$ are
used). The change in the vertical scale is due to charge creation (no
adjustment or renormalization has been made). }%
\label{superfig}%
\end{figure}

\subsection{Illustrations}

We show in Figs. \ref{superfig} and \ref{ultrafig} the time evolution of the
charge corresponding to the initial wavepacket (\ref{gau0}) in supercritical
wells. The only difference between both figures is the well depth,
$V_{0}=5mc^{2}$ in Fig. \ref{superfig} and $V_{0}=50mc^{2}$ in Fig.
\ref{ultrafig}. The calculations are semi-analytical in the sense that the
integration in Eq. (\ref{timeevol}) must be done numerically for each
space-time point $(t,x)$.

\begin{figure}[htb]
\includegraphics[scale=0.12]{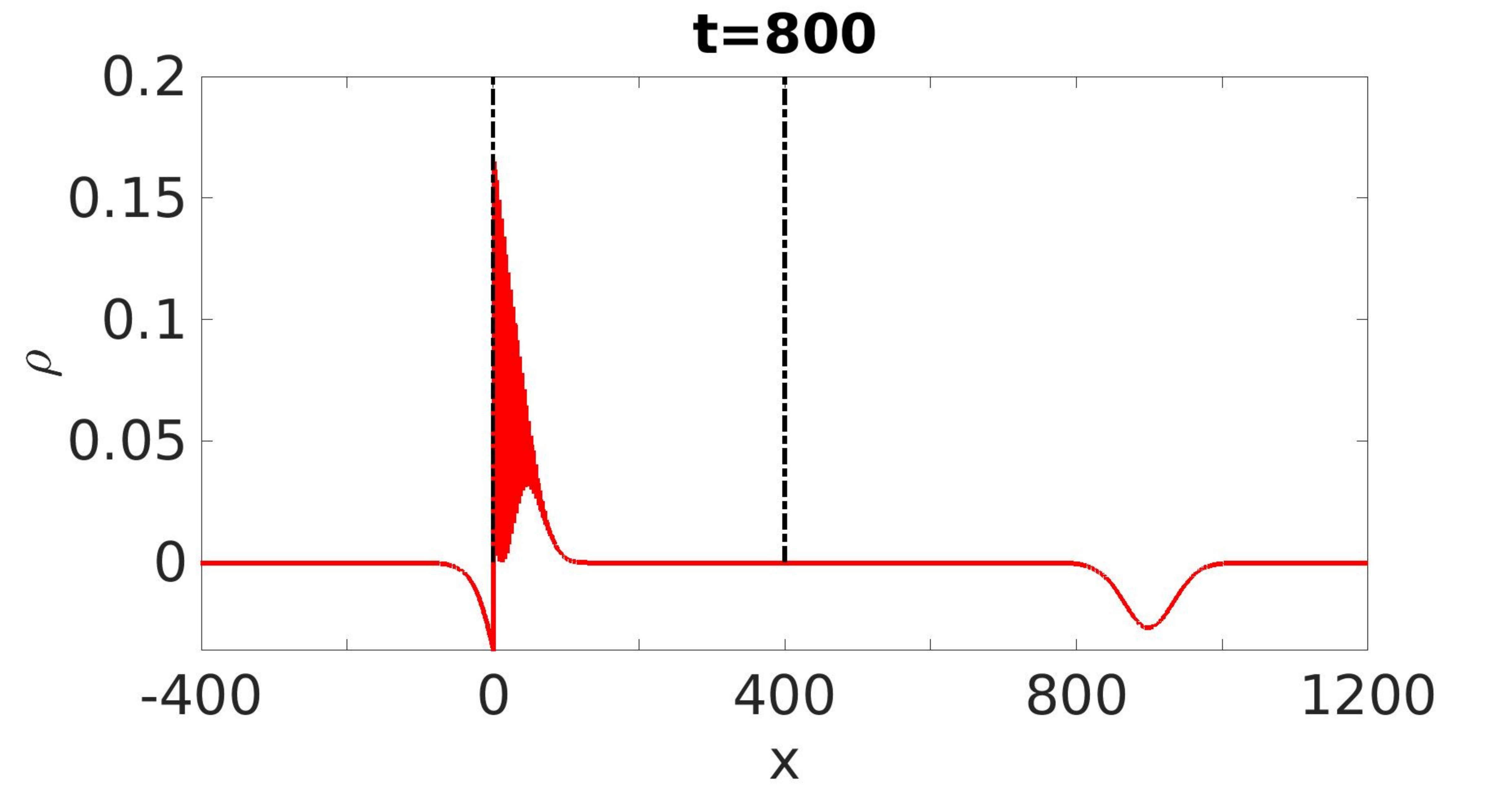}
\includegraphics[scale=0.12]{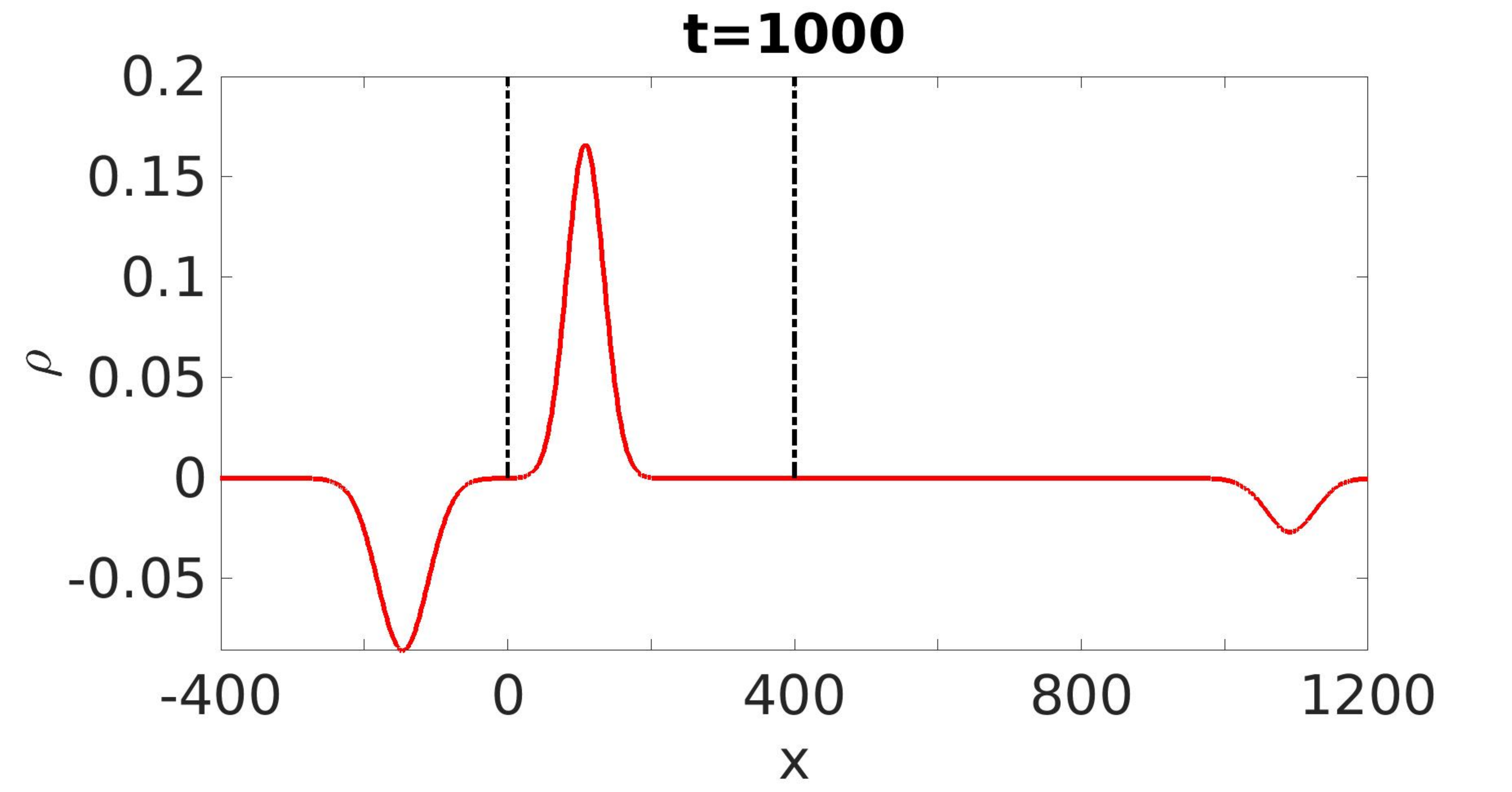} \caption{The charge density for
the system shown in Fig. \ref{superfig} as given by numerical computations
from a finite difference scheme (only the results at $t=800$ and $t=1000$ are
shown).}%
\label{superfdfig}%
\end{figure}

For $V_{0}=5mc^{2}$, Klein tunneling is prominent: the positive charged
wavepacket moves towards the right, and upon reaching the right edge, the
supercritical potential produces negative charge outside the well
(corresponding to antiparticles) and positive charge inside.\ The reflected
charge is higher than the incoming charge -- this is a time-dependent version of
Klein's paradox -- but
the total charge is conserved. The reflected wavepacket then reaches the left
edge of the well, resulting in a transmitted negatively charged wavepacket
 and a reflected wavepacket with a higher positive charge,
  now moving to the right inside the well. We have also
displayed (Fig. \ref{superfdfig}) results obtained from solving numerically
the KGE equation through a finite difference scheme. The numerical method
employed is described elsewhere \cite{paper2} -- here its use is aimed at
showing the accuracy of our MSE based wavepacket approach.

For a higher confining potential (Fig. \ref{ultrafig}), transmission outside
the well is considerably reduced: the wavepacket is essentially reflected
inside the well.\ This is due to the fact, noted above, that the plane-wave
transmission amplitudes from which the wavepacket is built are proportional to
$1/V_{0}$. Hence in the limit $V_{0} \rightarrow\infty$, Klein tunneling
becomes negligible. We recover a behavior similar to the one familiar for the
non-relativistic infinite well wavepackets \cite{robinett}.

\begin{figure}[htb]
\includegraphics[scale=0.12]{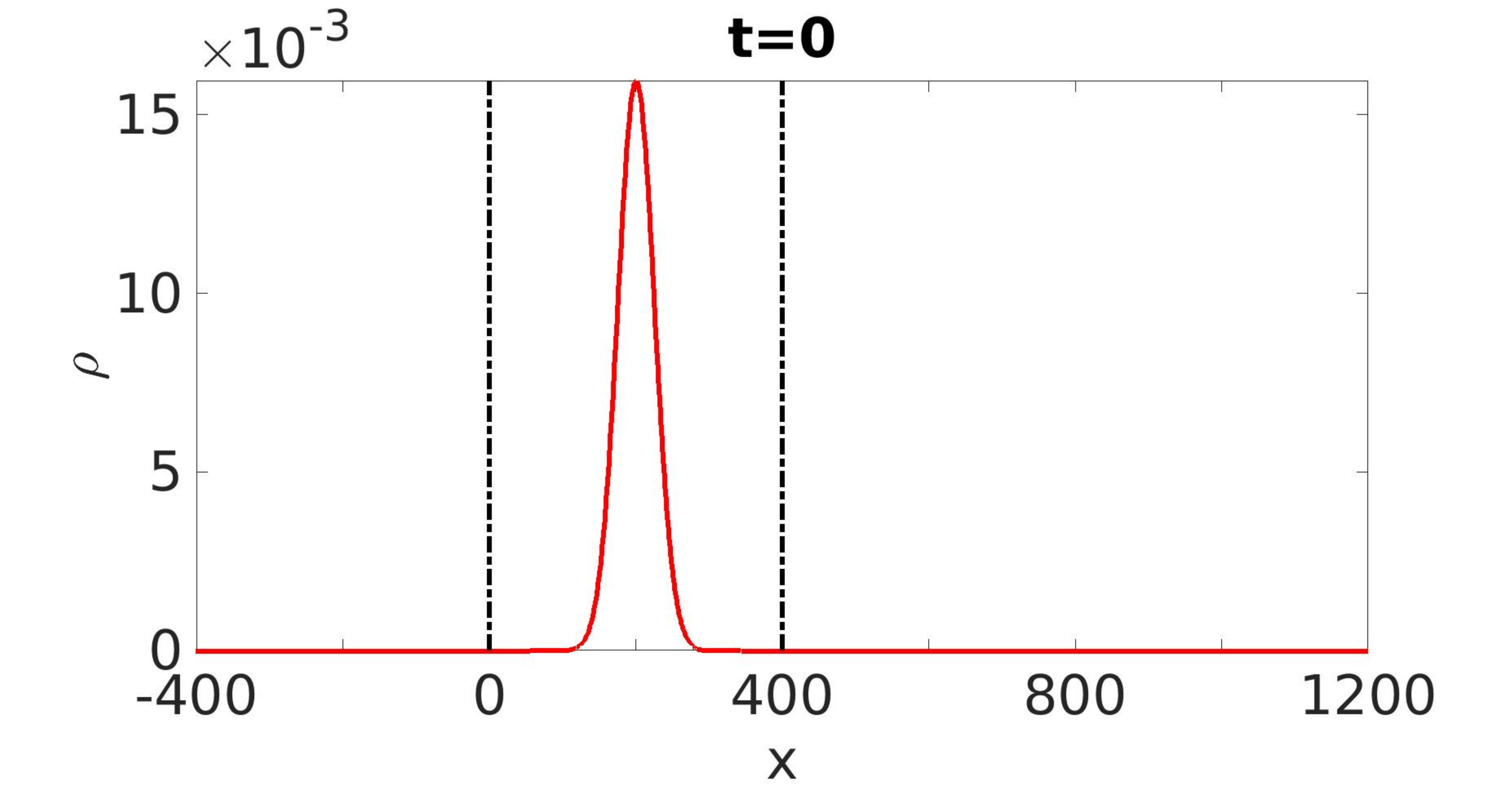}
\includegraphics[scale=0.12]{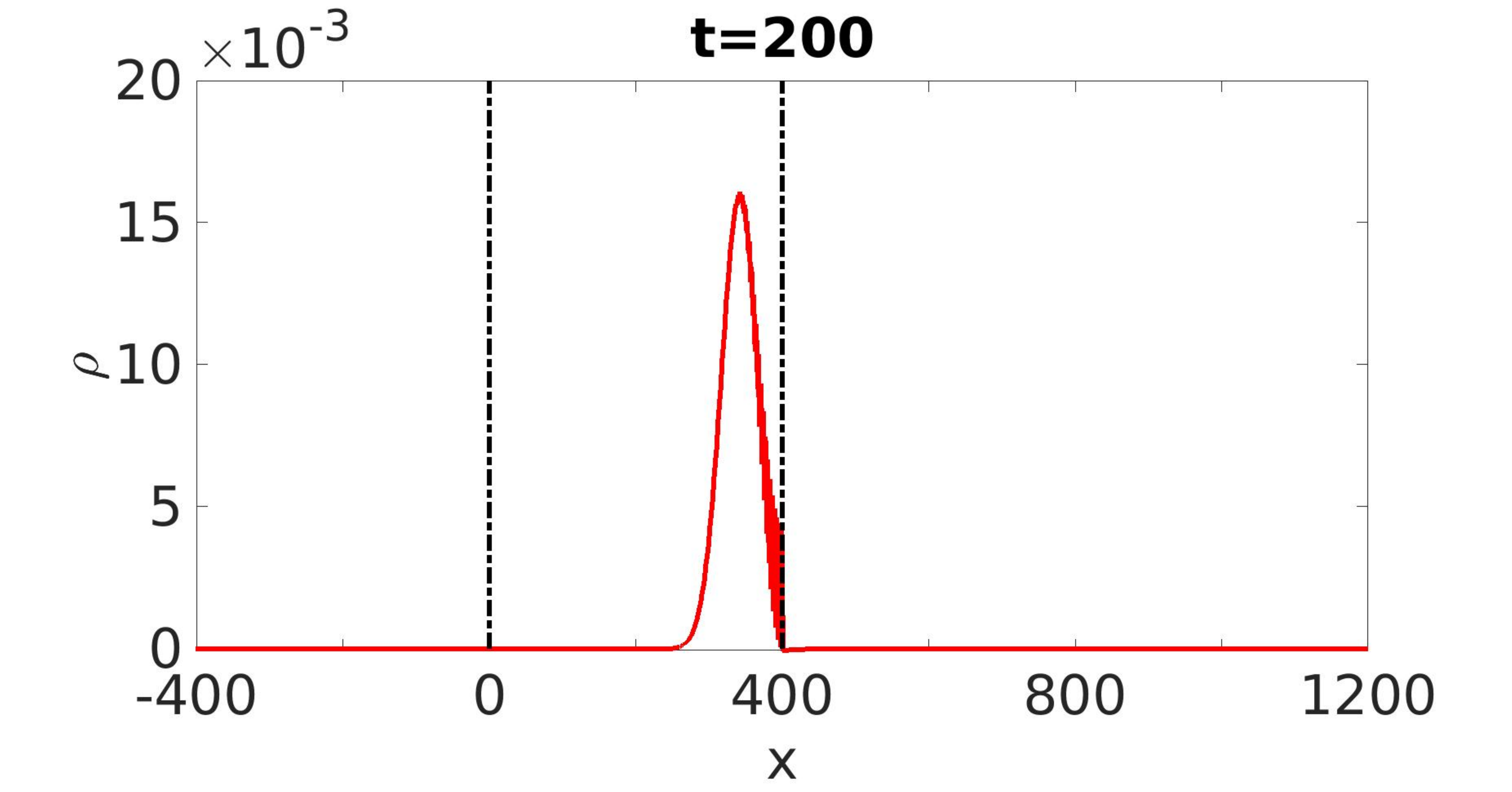}
\includegraphics[scale=0.12]{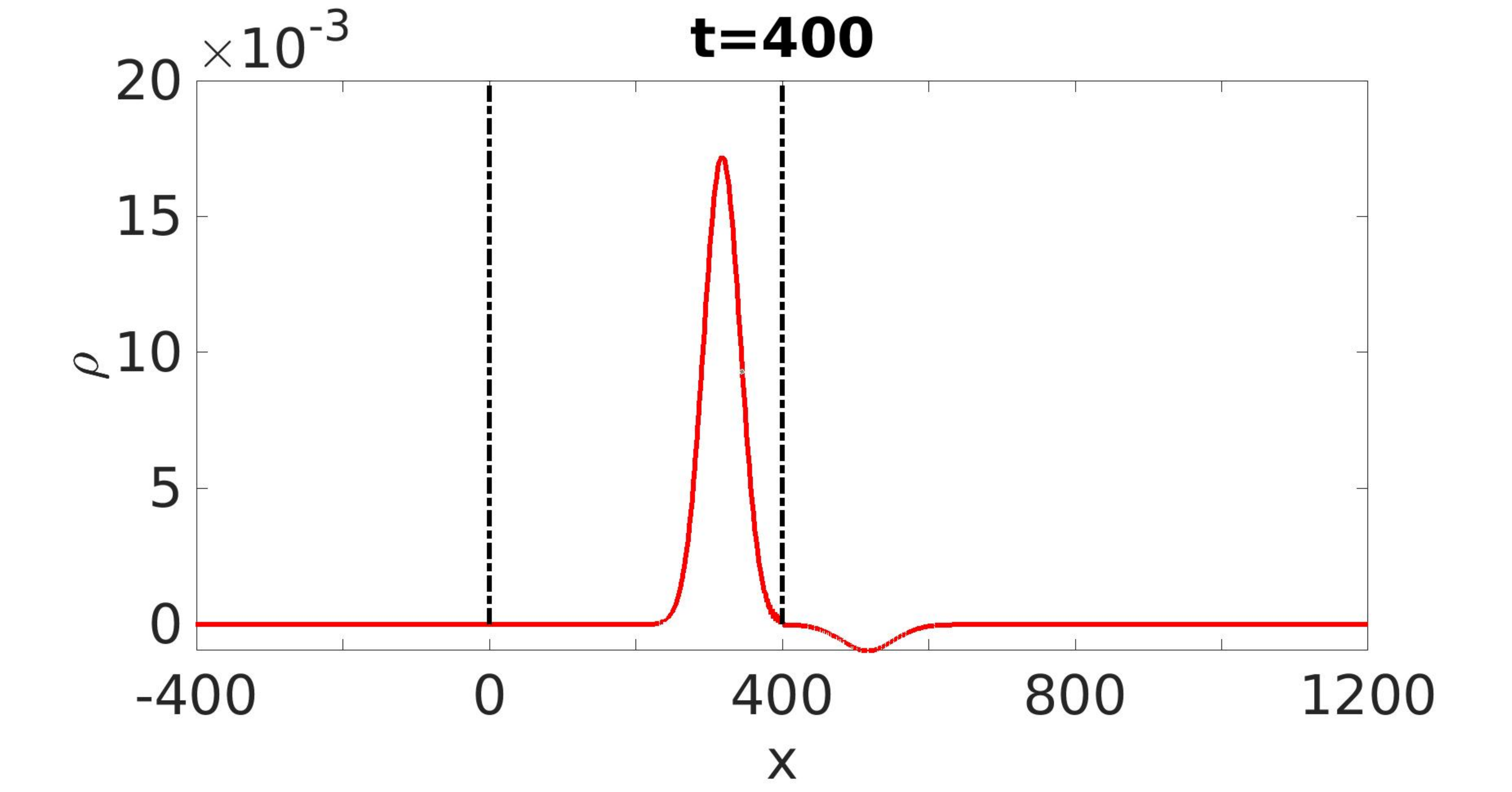}
\includegraphics[scale=0.12]{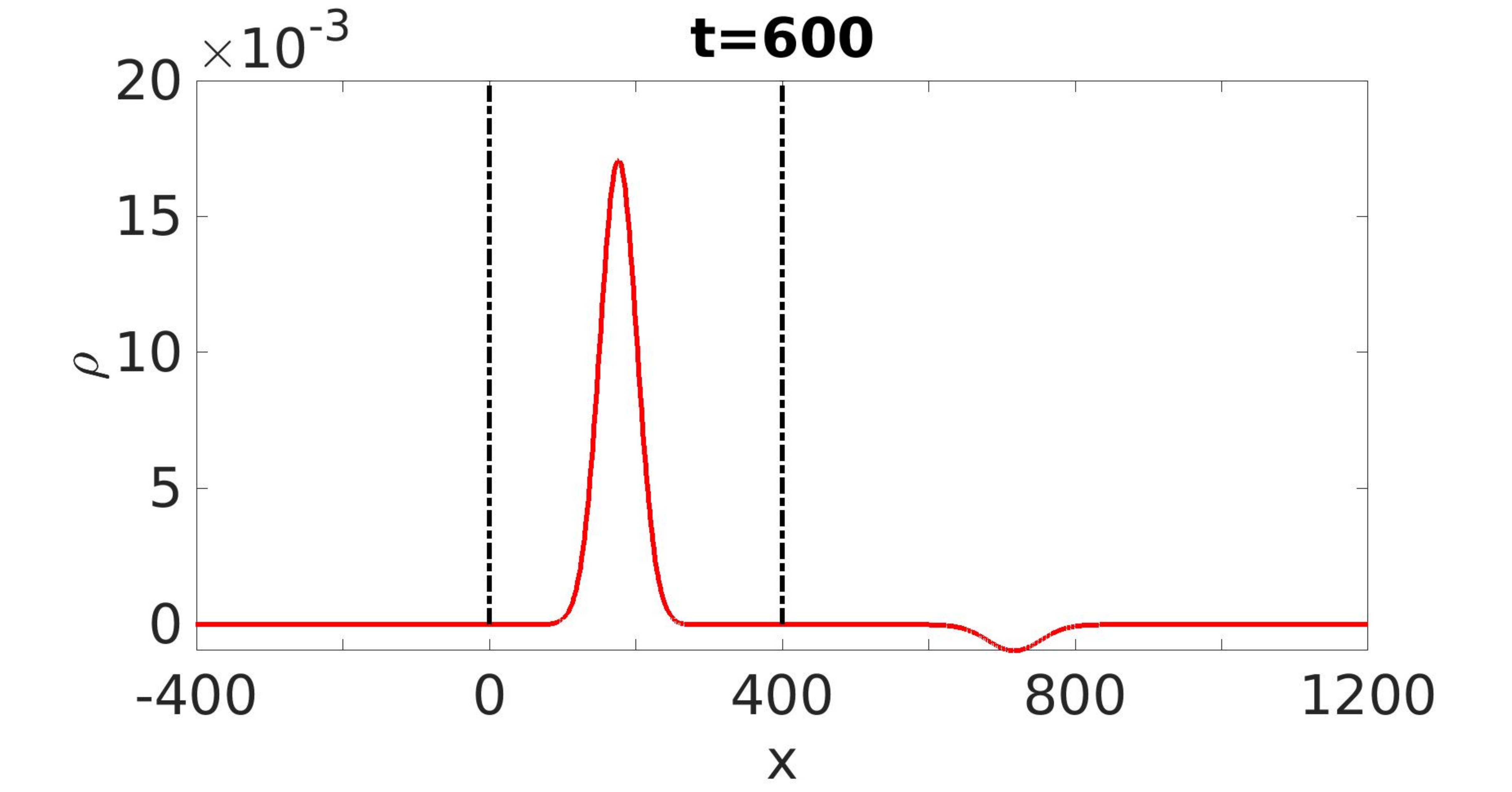}
\includegraphics[scale=0.12]{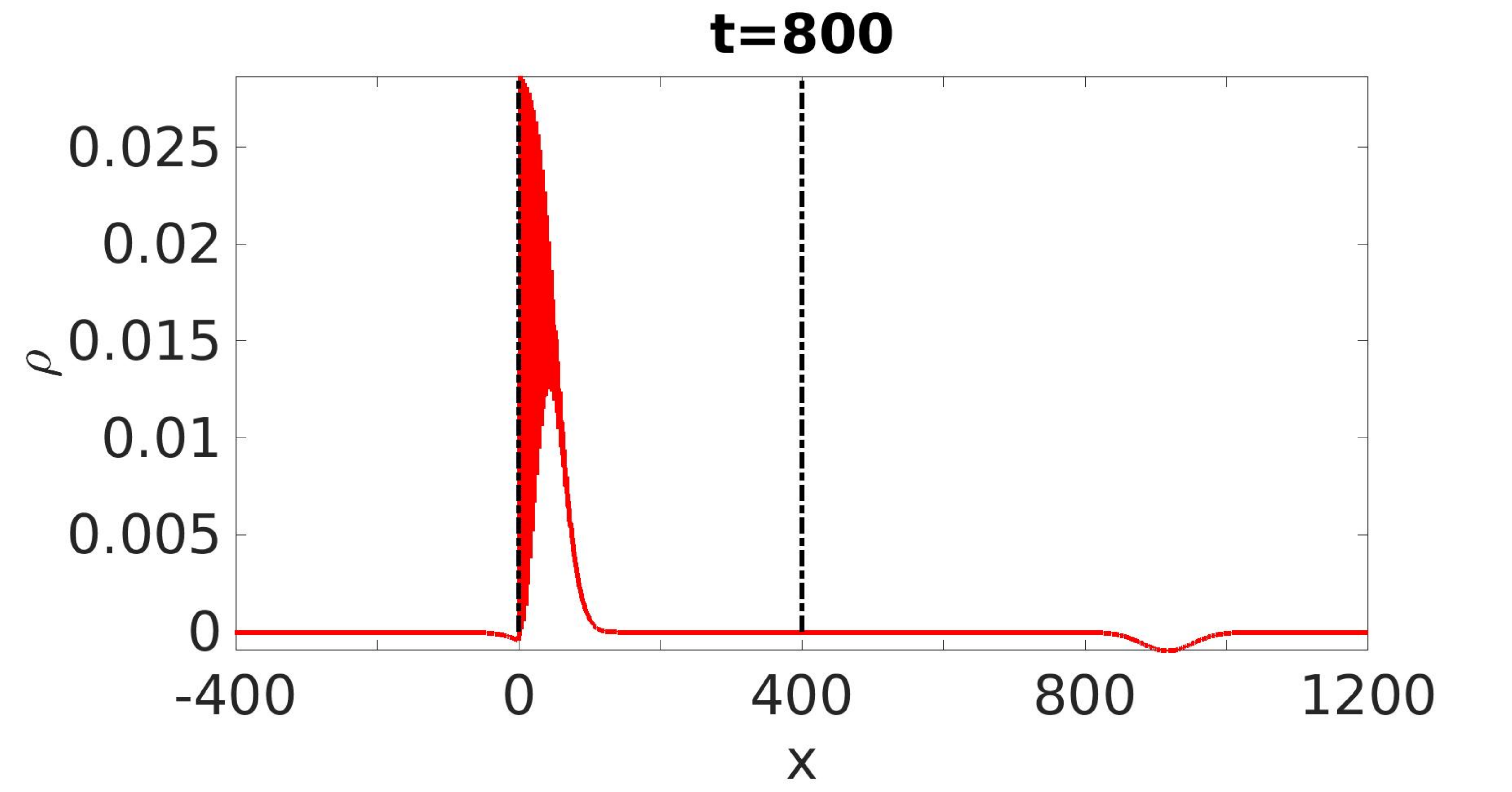}
\includegraphics[scale=0.12]{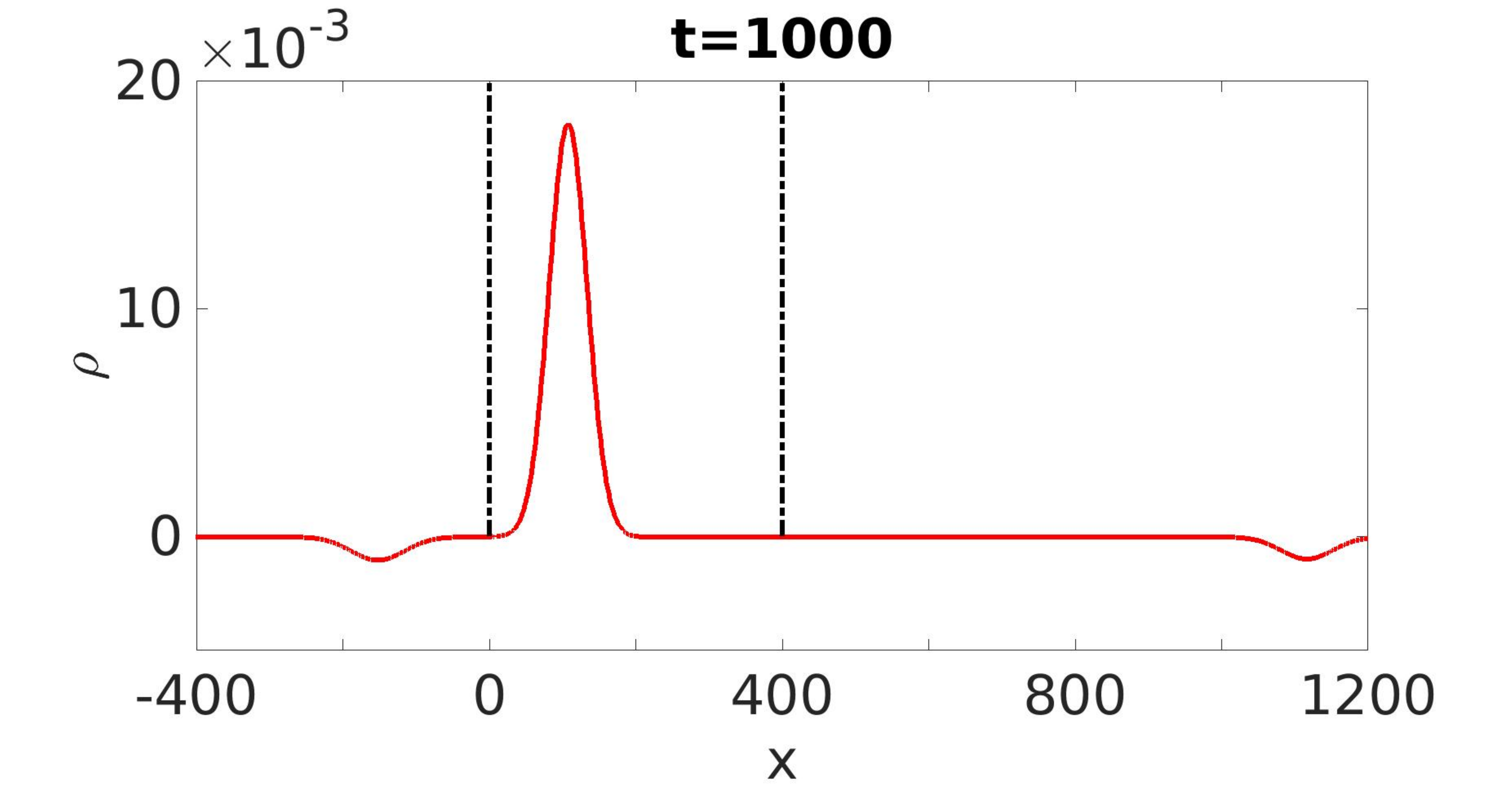} \caption{Same as Fig.
\ref{superfig} but for a well of depth $V_{0}=50$ $mc^{2}$. Klein tunneling is
suppressed relative to Fig. \ref{superfig}.}%
\label{ultrafig}%
\end{figure}

\section{Discussion and Conclusion}

In this work we studied a Klein-Gordon particle in a deep (supercritical)
square well. We have seen that the method based on connecting the
wave-function at both potential discontinuities, employed for non-relativistic
square wells, only works for non supercritical wells.\ For supercritical
wells, a divergent multiple scattering expansion was introduced to obtain the
solutions.\ This expansion accounts for Klein tunneling and for the Klein
paradox. In the limit of an infinitely deep well, the amplitudes obtained from
the expansion show that Klein tunneling is suppressed. The quantized particle
in a box similar to the non-relativistic one is then recovered, although
contrary to the non-relativistic case, this happens by oscillating Klein
tunneling solutions becoming negligible (rather than through exponentially
decaying wavefunctions becoming negligible outside the well). We have also
seen how these amplitudes can be used to build time-dependent wavepackets.

The methods employed here to study the square well for a relativistic spin-0
particle can be understood readily from the knowledge of non-relativistic
quantum mechanics. These methods have allowed us to introduce in a simple way
specific relativistic traits, such as charge creation (that in the
Klein-Gordon case already appears at the first quantized level) or Klein
tunneling and the Klein paradox. In particular, the wavepacket dynamics give
an intuitive understanding of these phenomena that are not very well tackled
in a stationary approach.

The framework employed in this paper -- that of relativistic quantum mechanics
(RQM) -- lies halfway between standard quantum mechanics and relativistic
quantum field theory (QFT). Indeed, RQM describes formally a single particle
wavefunction with a spacetime varying charge, while the physically correct
account afforded by QFT involves creation and annihilation of particles and
their respective antiparticles. The correspondence between the RQM and QFT descriptions for a
boson in the presence of a background supercritical potential has
been worked out in details \cite{gitman} for the case of the step potential
discussed in Sec. \ref{prin}. According to QFT, the potential spontaneously
produces particle/antiparticle pairs, a feature that is absent from the RQM
description.  For a Klein-Gordon particle, the RQM wavefunction correctly
represents the incoming boson as well as the QFT enhancement to the pair
production process; the enhancement results from the interaction between the incoming boson
and the supercritical potential (this is the charge increase visible in Fig.
\ref{superfig}). This correspondence can be established in a time-independent
approach \cite{manogue,holstein}, or more conclusively by employing space-time
resolved QFT calculations \cite{grobe-boson}. From an experimental viewpoint, 
direct pair production from a supercritical background
field has remained elusive up to now, though the current development of strong laser facilities 
could lead to an experimental observation (for the fermionic electron-positron pair production) in a foreseeable future \cite{reviewE}. The bosonic supercritical well and the conditions under which 
quantized energy levels could be observed is not at present experimentally on the table.

Note finally that the disappearance of Klein tunneling in the infinite well
limit should be of interest to recent works that have studied the Klein-Gordon
equation in a box with moving walls \cite{koehn,hamidi,colin} (the special
boundary conditions chosen in these works were indeed not justified). The
method employed here for spin-0 particles obeying the Klein-Gordon equation is
also suited to treat a spin-1/2 particle in a square well obeying the Dirac
equation. The scattering amplitudes in the Dirac case will however be
different, and the results obtained here for spin-0 particles regarding the
suppression of Klein tunneling in infinite wells will not hold.

\end{document}